# High-current CNT films grown directly on commercially available 2.5D substrates for low-voltage field-emission electron sources

Nannan Li[a,b], Fei Yan[b], Baoqing Zeng[c], and Yi Luo[a,*]

[a]Micro/Nano Fabrication Laboratory, Microsystem & Terahertz Research Center, China Academy of Engineering Physics (CAEP), Chengdu, 610200, China

[b]Institute of Electronic Engineering, CAEP, Mianyang, 621900, China

[c]National Key Laboratory of Science and Technology on Vacuum Electronics, University of Electronic Science and Technology of China, Chengdu 610054, China

*E-mail: luoyi@mtrc.ac.cn

Abstract

Carbon nanotube (CNT) based electronic devices are promising for beyond-silicon solid-state electronics and vacuum micro-nano-electronics. Despite rapid progress in CNT field-effect transistor related solid-state electronics, the development of CNT-based vacuum nanoelectronic devices is substantially blocked by the longstanding challenges in demanding high-current field-emission (FE) electron sources at low operating voltage. In addition to CNTs' properties, FE characteristics are also affected by substrate morphology and interface state. This work demonstrates high-current FE characteristics at relatively low operating voltage by using CNT films grown directly on commercially available 2.5D substrates with matched feature size and improved interface contact. Simulation results indicate that the commercially available 2.5D substrate including nickel foam (NiF) and carbon cloth (CC) with appropriate feature size would dramatically help to enhance emission current at a relatively lower voltage. Modified fabrication process results in improved contact between CNTs and the underlying 2.5D substrates. Twenty times higher emission current density with halved lower turn-on electric field achieved by CNTs grown directly on randomly picked NiF shows the potential of 2.5D substrate with good contact in improving FE characteristics. Finally, a high emission current (6 mA) with approximately 75% decrease in turn-on electric field was realized by matching the feature size of 2.5D substrate with that of CNTs, bringing us significantly closer to reliable high-current and low-voltage FE electron sources for practical applications.

1. Introduction

Owing to their extraordinary physical, chemical, and mechanical properties, carbon nanotubes (CNTs) [1] have attracted tremendous interests in future electronic systems consisting of solid-state electronics and vacuum micro-nano-electronics. Recently, there have been dramatic breakthroughs in CNTs-based field-effect transistor (CNFET) research towards beyond-silicon solid-state electronics [2-4]. However, the development of CNTs-related vacuum micro-nano-electronics is longstanding blocked by technological issues from the "heart" field-emission (FE) electron sources. One of the most serious challenges is that high current and low operating voltage could not be achieved simultaneously. Overcoming these bottlenecks is of great significance for enabling fully extended CNTs-based electronic applications.

CNTs are attractive FE electron source candidates due to their exceptionally high aspect ratio, excellent mechanical strength, outstanding chemical stability, and remarkable conductivities [5-8]. The practical applications in vacuum micro-nano-electronics, such as compact X-ray sources [9,10], new-generation microwave tubes [11, 12], and miniature electron guns [13], generally require a high current typically in mA range at a relatively low operating voltage [14,15]. CNT emitters with substantial electron emission area, falling into two categories: vertically aligned CNT (VACNT) arrays and CNT films, are needed. VACNT arrays with optimal inter-CNT separations show excellent low-voltage capabilities [16], however, emission current is limited by the emission degradation and the relative short lifetime derived from the ion-sputtering/heating destruction of those dominating tips [17, 18]. Thus VACNT arrays have been seldom employed as practical electron sources. By contrast, CNT films could be used as a robust FE electron sources which combine long lifetime, relative ease of fabrication, and possibility to scale for production [19]. Tremendous progresses [20-24] have made on enhancing emission current of CNT films, resulting in high current in the range of several to tens of milliamperes. However, the future practical application is seriously limited by the high turn-on electric field typically in the range of 4-10 V/μm due to the suppressed field enhancement and increased field screening effect. To reduce turn-on electric field of

CNT films, a widely-reported technique is to construct CNTs-based multistage structures such as CNTs on Si tips [25], CNTs on ZnO nanowire arrays [26], and CNTs on metal cones [27], which however induce additional drawbacks from the 3D multistage structures fabrication and repeatability. Recently, CNT films on commercially available 2.5D substrates consisting of cross-linked horizontal 1D microstructures such as nickel foam (NiF) [28,29] and carbon cloth (CC) [30-32] offered a robust alternative for encouraging low-voltage FE devices. Significantly improved turn-on electric field in the range of 0.4-0.8 V/μm has been achieved, promising practical applications in field emission displays (FEDs). Nevertheless, their typically small emission current (< 1 mA) would block most of applications in vacuum micro-nano-electronics. The FE characteristics are complicated results of many factors, which depend on CNTs' properties and also affected by substrate morphology and interface contact. CNTs inherently promise excellent FE characteristics owing to their extraordinary physical, chemical, and mechanical properties. The practical FE performance is generally limited by the substrate and interface properties. Therefore, there is a pressing need to understand the influence mechanism and develop an accessible high-current CNTs films with relatively low turn-on electric field for future FE electron sources.

   Herein, we demonstrate high-current (~ 6 mA) CNT films grown directly on commercially available 2.5D substrates with a low turn-on electric field (~ 0.55 V/μm) by matching the feature size of 2.5D substrate with CNTs and improving the interface contact between CNTs and the substrate, promising CNT-based FE electron sources for practical applications. This is the first time to report high emission current and low turn-on electric field simultaneously for CNTs films on a commercially available 2.5D substrate. Simulation investigations indicate that the 2.5D substrate with appropriate feature size would dramatically enhance emission current and decrease working voltage. This finding facilitates the realization of high-current and low-voltage FE devices by using an appropriate commercially available 2.5D substrate to support CNT emitters. Experimentally, CNT films were grown directly on 2.5D substrates without catalyst deposition to improve the adhesion and contact resistance between emitters and substrate, which is benefit to enhance high-current delivery. Twenty times higher current density with 50% decrease in the operating voltage was demonstrated by CNTs directly grown on randomly picked NiF compared with that of CNTs on flat silicon. An optimized 2.5D substrate, e.g. CC, with feature size comparable to CNT height was used to support the directly grown CNT films, achieving a much higher FE current ( ~6 mA) with approximately 75% decrease in turn-on electric field. The much better FE characteristics are attributed to the maximum field enhancement induced by the 2.5D substrate and the improved adhesion and contact between CNTs and substrate. The high emission current is adequate to enable practical FE electron source in vacuum micro-nano-electronics, such as compact X-ray sources, new-generation microwave tubes, and miniature electron guns.

2. Experimental

CNTs were synthesized on different substrates via thermal chemical vapor deposition (TCVD) technique, which presents a potential solution for low-cost, atmospheric-pressure and large-scale synthesis of high-quality CNT films on irregular-shaped substrates. Compared with the screening-printing CNTs on 2.5D substrates [28,29], the directly grown CNTs exhibit improved adhesion and contact resistance, which is beneficial for electron transport and current collection. The TCVD synthesis of CNTs was implemented by using the system with a horizontal corundum tube placed in the furnace. First, the substrates were ultrasonic cleaned in acetone alcohol de-ionized water respectively, and dried by nitrogen gas flow to avoid pollutants. Then, a nickel-aluminium-nickel multilayer serving as catalytic was deposited on Si substrates which were used as references. By contrast, 2.5D substrates were treated by specific process without direct deposition of catalyst to further improve the contact interface between CNTs and underlying 2.5D substrate. NiF was activated in acid solution since the nickel contained in NiF can act as the catalyst for CNT growth. CC was immersed in 0.5 M ferric sulfate heptahydrate solution to be activated. Subsequently, one of these substrates was placed in a small ceramic boat lying in the middle of the corundum tube. Both the end of the corundum tube was sealed with gas pipelines. Hydrogen at a flow rate of 135 sccm was introduced to expel the residual air in the corundum tube. The furnace was then heated to 750 $^{\circ}$C in hydrogen. Acetylene with a flow rate of 50 sccm was introduced and carried by hydrogen to pass over the substrates for 0.5 h. Hydrogen was co-flowed to systematically achieve a suitable balance for synthesis by neutralizing excess reactive carbon species and opposing the decomposition of the hydrocarbon feedstock, resulting in high-quality CNTs with very little amorphous carbon present. The reactor was cooled down to room temperature in ambient hydrogen. After reaction, there was a layer of CNTs deposited on the substrates. The density and length of CNTs can be controlled by adjusting the growth time. The growth conditions were refined to ensure the CNTs with high crystallizability and no amorphous carbon layer at interface between CNTs and substrate, facilitating the improvement of contact and adhesion at the interface. This is beneficial for high-current transport and collection.

Surface morphology of CNTs' films on different substrates was characterized by scanning electron microscopy (SEM). The FE behaviors were measured using a classical parallel-diode configuration in an ultrahigh-vacuum chamber. The CNT emitters on different substrates were loaded on a stainless-steel holder that incorporates the emitters as the base electrode. A parallel nickel disk with diameter of 40 mm served as the anode at a fixed base-anode spacing of 600 μm. Both holder and anode were undergone heat treatment to remove contaminants and degassing before mounting. Test was performed in an ion-pumped vacuum chamber reaching a base pressure of $10^{-8}$ Torr after a standard vacuum bakeout for outgassing.

3. Results and discussion

Field emission behaviors depend on the local electric field on CNT films. In this work, the enhanced local electric field around the CNTs' tip on commercially available 2.5D substrate consisting of cross-linked horizontal 1D microstructures were investigated by electrostatic simulation using finite integration technique [33]. Fig. 1a shows the schematic of individual CNT on a horizontal cylinder. Compared with the reported rectangular model [34] and hemi-ellipsoid tip [28,35], the lying cylinder is more approaching the actual geometry of the cross-linked lying 1D structures in the

2.5D substrates. Simulation was carried out to extract the potential and electric field at any point around the structure. Fig. 1b shows the potential contour distributions for CNT on flat and 2.5D substrate, respectively. The more concentrated potential lines around the CNT on cylinder indicate stronger local electric field, resulting in high emission current at relatively low operating voltage. This qualitatively demonstrates that a randomly picked 2.5D substrate would help to enhance emission current and decrease working voltage.

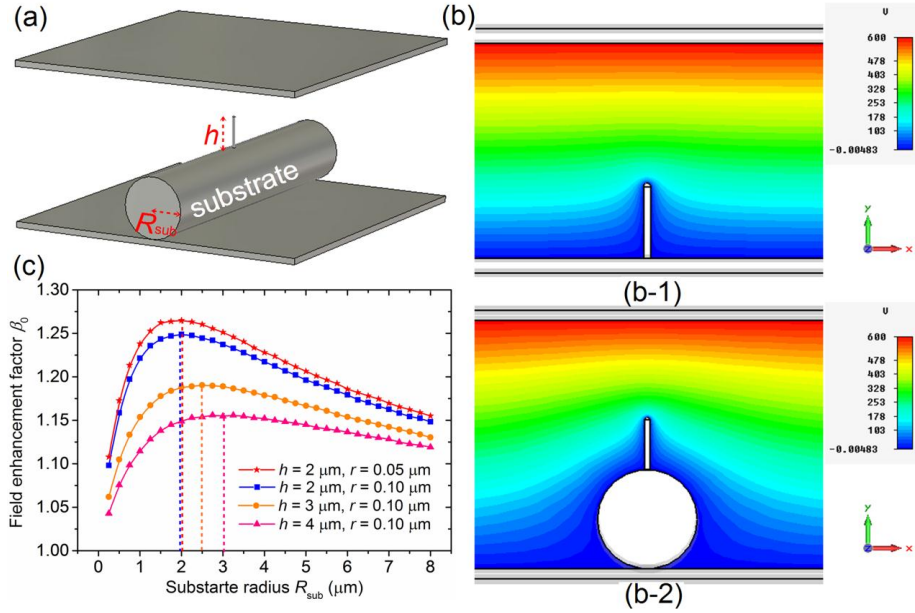

**Fig. 1.** Electrostatic simulation results of enhanced local electric field around CNT on 2.5D substrate. (a) Model of CNT on 2.5D substrate. (b) Potential contours of CNT on (b-2) 2.5D substrate compared with that (b-1) on flat substrate. (c) Field-enhancement factor $\beta_0$ induced by the 2.5D substrate with various feature size $R_{sub}$ for a series of CNTs with various height $h$ and tip radius $r$.

To quantificationally explore the effect of 2.5D substrate on improving FE characteristics of CNTs, we assume the local electric field at the apex of CNT on flat substrate and on a lying 1D microstructure is $E_{ref}$ and $E_{apex}$ respectively. The field-enhancement factor induced by the 2.D substrate can be expressed as $\beta_0 = E_{apex} / E_{ref}$. To explore the gain in emission current, the ratio of emission current for sample *b-2* and *b-1* shown in Fig. 1b under the same applied voltage can be obtained according to Fowler-Nordheim (FN) theory [36, 37]:

$$I_{b-2} / I_{b-1} = \beta_0^2 \exp[0.95B\phi^{3/2}(1-1/\beta_0)/E_{ref}] \quad (1)$$

where $B = 6.83089$ eV$^{-3/2}$V/nm and $\phi$ is the effective work function of CNT. It is worth noting that $\beta_0$ would not only cause the operating voltage decrease but also bring emission current enhancement. Here, the relationship of $\beta_0$ with the curvature radius of underlying cylinder $R_{sub}$ is investigated for a series of CNTs with different height $h$ and tip radius $r$, as shown in Fig. 1c. When $R_{sub}$ is infinitely large, the underlying cylinder could be treated as a flat substrate. Thus, $\beta_0$ would approach to 1. When $R_{sub}$ is small enough (for example, close to or smaller than the tip radius of CNT $r$), the field enhancement induced by the underlying cylinder is also negligible. Simulation results suggest that there is an optimized $R_{sub}$ to enable the highest $\beta_0$ for a certain CNT. The optimized $R_{sub}$ increases slowly with CNTs' height $h$. The maximum $\beta_0$ depends on the aspect ratio ($h/r$) of individual CNT. Therefore, a 2.5D substrate is beneficial to enhance emission current and decrease working voltage due to the enlarged electric field strength around the CNTs' tip, and the 2.5D substrate with matched feature size with that of CNTs would dramatically enhance emission current and decrease working voltage due to the maximum $\beta_0$. These findings provide important implications for tailoring high-current CNTs-based FE devices for various applications by utilizing an appropriate commercially available 2.5D substrate.

In addition to the geometrical morphology of substrate, the substrate material and contact interface properties are crucial. NiF has being considered desirable 2.5D substrate for supporting CNTs due to the excellent electrical and thermal conductivities as well as relatively good adhesion between CNT films and substrates. A comparative study of FE characteristics for CNT films deposited at flat Si and NiF substrates has been performed and shown in Fig. 2. Fig. 2a and 2b show the measured FE characteristics of CNT films on flat Si, which is considered as references. An emission current density of 1 mA/cm$^2$ requires an electric field of 4.4 V/μm. The turn-on electric field ($E_{on}$) is around 2.3 V/μm. The approximately linear behavior shown in Fig. 2b indicates that the electrons were mainly extracted through quantum tunneling FN mechanism. Field-enhancement factor $\beta$ of CNT films estimated from the slope of FN plot is approximately 4400, which strongly depends on the surface morphology of the CNT films. By contrast, Figs. 2c and 2d show the measured FE properties of the CNT films on NiF fabricated under the same conditions. An approximately 20 mA/cm² high emission current density was achieved when applying an electric field of 4.4 V/μm. The turn-on electric field ($E_{on}$) is around 1.1 V/μm, which is about 50% smaller than that of CNTs on flat Si. The field enhancement factor of CNTs-NiF cathode derived from the FN plot shown in Fig. 2d is around 9400, which is the result of geometrical configuration of CNTs on NiF. CNT films on NiF exhibits a twenty times improvement in FE current density, which should be mainly attributed to the field enhancement induced by the NiF and the improved adhesion and contact resistance between the directly grown CNTs films and the NiF substrate. The emission current density is higher than that of currently reported screen-printing CNTs on NiF [28,29] and that of recently reported VACNTs [38,39].

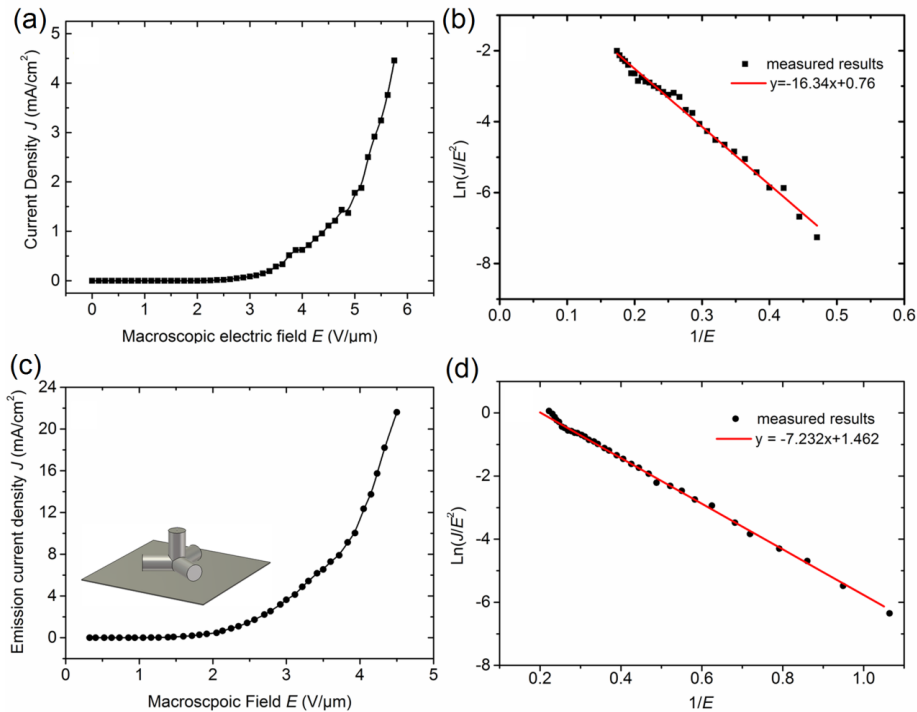

**Fig. 2.** Experimentally measured field-emission characteristics of CNT films directly grown on flat Si and 2.5D NiF substrates respectively under same conditions. (a) *J-E* curve and (b) FN plot of CNTs films on Si. (c) *J-E* curve and (d) FN plot of CNTs films on NiF.

Despite improvement in FE characteristics for CNT films directly grown on NiF, NiF is not an ideal 2.5D substrate for supporting CNT films owing to the mismatch of feature size. Figure 3a-c show the SEM images of CNT films on NiF at different magnification. The curvature radius of the lying branches of NiF is around 100 μm, which is much larger than the CNT length. Thus, the field-enhancement factor $β_0$ induced by the lying horizontal 1D microstructure is approaching 1. The standing branches on the lying ones as shown in Fig. 3b, which is uncontrollable and randomly distributed, dominate the field enhancement effect of the NiF substrate. To maximize the field enhancement effect of 2.5D substrate, CC consisting of woven carbon fibers is a promising substrate since the curvature radius of the carbon fibers could be matched with the CNT size as suggested by the simulation results. However, the emission current is generally limited by the poor contact properties between CNTs and CC substrate. Here, significant improvement in mechanical adhesion and contact resistance between CNTs and CC substrate has been achieved by the strong C-C bond at the interface derived from the direct growing CNTs on CC without catalyst deposition. The contact properties have been further improved by the no amorphous carbon layer in the interface derived from the refined growth conditions. Figure 3d-f show the SEM images of CNT films on CC at different magnification. The CNTs grown directly on the CC with feature size comparable to CNT height and improved interface properties are promising to achieve much higher FE current at relatively low operating voltage.

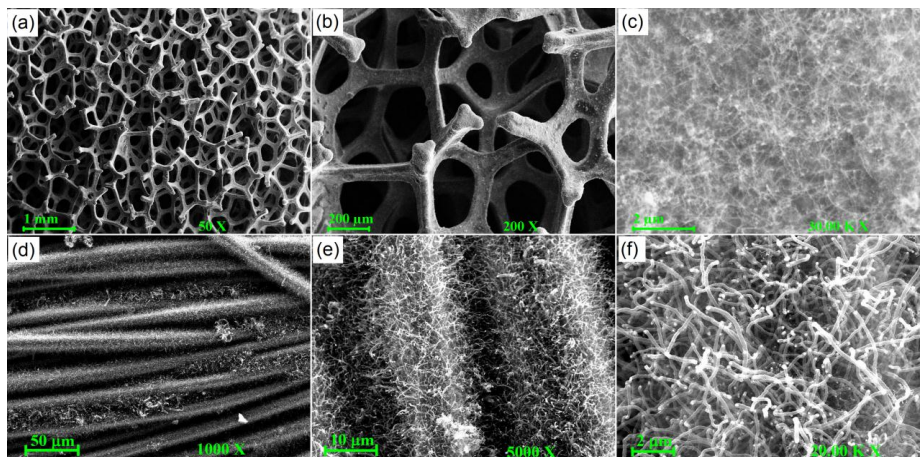

**Fig. 3.** SEM images of CNT films grown directly on NiF and CC under same conditions. (a) CNTs on NiF; (b) CNTs on NiF at high magnification; (c) CNTs on NiF at higher magnification. (d) CNTs on CC; (e) CNTs on CC at high magnification; (f) CNTs on CC at higher magnification.

The remarkable FE characteristics of CNTs on CC are shown in Fig. 4. A higher than 6-mA emission current (~764 mA/cm$^2$) has been achieved on an emitting area with diameter of 1mm when applying an electric field of 4.4 V/μm. The high emission current is adequate to enable practical applications in vacuum micro-nano-electronics, such as compact X-ray sources, new-generation microwave tubes, and miniature electron guns. Simultaneously, a low $E_{on}$ around 0.55 V/μm, which is 75% smaller than that of CNT on Si, was also achieved. It is worth noted that the CNTs on different substrates were fabricated under same conditions. The different morphology of CNTs on NiF and CC, as shown in Fig. 3c and Fig. 3f, should be attributed to the substrate effect. The CNTs on NiF show much higher aspect ratio (*h/r*) than that on CC, which should lead a much larger emission current and a relatively lower working voltage regardless of the field enhancement effect induced by substrates. Therefore, the much better FE characteristics from CNTs on CC should be mainly attributed to the CC substrate. Compared with the recent reports [28-32] which focused on low turn-on electric field at small emission current density, we simultaneously achieved high emission current and low turn-on electric field by using CNTs films directly grown on commercially available 2.5D substrate for the first time.

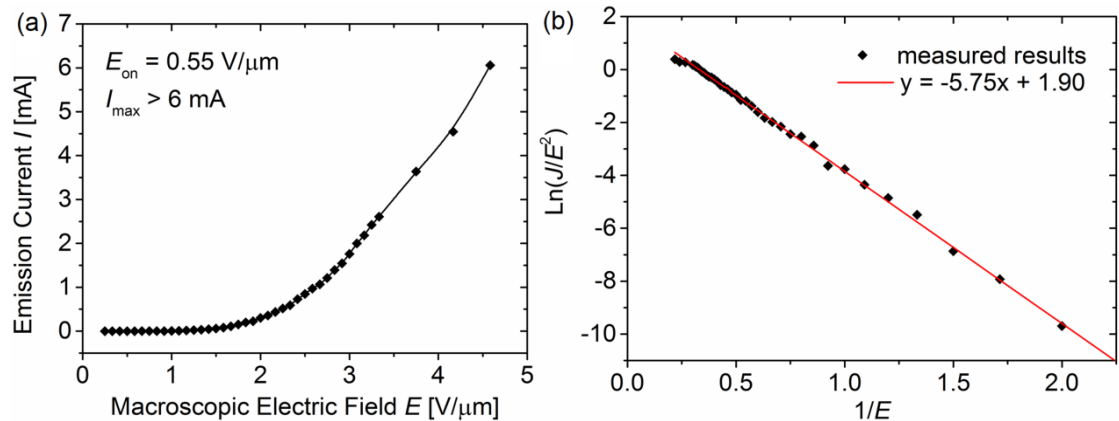

**Fig. 4.** Excellent field-emission characteristics of CNTs on CC. (a) *I-E* plot; (b) FN relationship.

In addition, 2.5D substrates (including NiF and CC) have several other advantages. First, their high electrical conductivity will help electrons transport. Second, the large specific surface area and high thermal conductivity are beneficial to heat dissipation, greatly preventing CNTs destruction at high temperatures. Large specific surface area will accommodate more emitters and decrease screening effects. Furthermore, the inherent flexibility of 2.5D substrates could extend the application area of field-emission devices into flexible electronics. It is worth noting that the FE performance of CNTs on 2.5D substrate would be further improved when the fabricated CNTs films are optimized and the physical and chemical properties of the 2.5D substrates are fully utilized.

4. Conclusions

In conclusion, this work demonstrates that an excellent high-current and low-voltage FE electron source could be achieved conveniently by CNTs films directly grown on commercially available 2.5D substrate when the feature size of 2.5D substrate matches with that of CNTs and the interface properties between CNTs and substrate are improved. The field enhancement effect from 2.5D substrates consisting of cross-linked horizontal 1D microstructures was investigated to reveal the high-current and low-voltage mechanism. The interface properties including mechanical adhesion and contact resistance, which are crucial to deliver high emission current, have been improved by modified fabrication process. An appropriate 2.5 substrate, e.g. CC, with feature size matched to that of CNT height and with excellent interface contact is conducive to enhance the field-emission performance from CNT emitters. The additional properties of the 2.5D substrates, including excellent electron transmission, alleviated electrostatic shielding effect, and the enhanced CNTs accommodation should be also taken into account for maximum emission current from CNTs. Herein, we simultaneously achieved both high emission current and low turn-on electric field for CNTs films on a commercially available 2.5D substrate, promoting practical low-voltage FE applications in future in vacuum micro-nano-electronics. This work is of great significance for enabling fully extended CNTs-based electronic applications.

Acknowledgments

This work was supported by the National Natural Science Foundation of China [grant number 61801449].